\documentclass[twocolumn,prb,amsmath,amssymb,superscriptaddress]{revtex4}
\usepackage{graphicx}% Include figure files
\usepackage{dcolumn}% Align table columns on decimal point
\usepackage{bm}% bold math
\usepackage{xcolor}

\begin{document}
\title{Fermi resonance in the Raman spectrum of graphene}

\author{Dipankar Kalita}\affiliation{Univ. Grenoble Alpes, Inst NEEL, F-38042 Grenoble, France} \affiliation{CNRS, Inst NEEL, F-38042 Grenoble, France}
\author{Michele Amato}\affiliation{Universit\'e Paris-Saclay, CNRS, Laboratoire de Physique des Solides, 91405, Orsay,  France}
\author{Alexandre Artaud}\affiliation{Univ. Grenoble Alpes, Inst NEEL, F-38042 Grenoble, France} \affiliation{CNRS, Inst NEEL, F-38042 Grenoble, France}\affiliation{CEA, INAC-PHELIQS, F-38000 Grenoble, France}
\author{La\"etitia Marty}\affiliation{Univ. Grenoble Alpes, Inst NEEL, F-38042 Grenoble, France}\affiliation{CNRS, Inst NEEL, F-38042 Grenoble, France}
\author{Vincent Bouchiat}\affiliation{Univ. Grenoble Alpes, Inst NEEL, F-38042 Grenoble, France} \affiliation{CNRS, Inst NEEL, F-38042 Grenoble, France}
\author{Johann Coraux}\affiliation{Univ. Grenoble Alpes, Inst NEEL, F-38042 Grenoble, France} \affiliation{CNRS, Inst NEEL, F-38042 Grenoble, France}
\author{Christian Brouder}\affiliation{Sorbonne Universit\'e, CNRS UMR 7590, MNHN, IMPMC, F-75005 Paris, France}
\author{Michele Lazzeri}\affiliation{Sorbonne Universit\'e, CNRS UMR 7590, MNHN, IMPMC, F-75005 Paris, France}
\email{michele.lazzeri@sorbonne-universite.fr}
\author{Nedjma Bendiab}\affiliation{Univ. Grenoble Alpes, Inst NEEL, F-38042 Grenoble, France} \affiliation{CNRS, Inst NEEL, F-38042 Grenoble, France}
\email{nedjma.bendiab@neel.cnrs.fr}

\date{\today}

\begin{abstract}

We report the observation of an intense anomalous peak at 1608 cm$^{-1}$ in the Raman spectrum of graphene associated to the presence of chromium nanoparticles in contact with graphene.
Bombardment with an electron beam demonstrates that this peak is distinct from the well studied D$'$ peak appearing as defects are created in graphene; the new peak is found non dispersive.
We argue that the bonding of chromium atoms with carbon atoms softens the out-of-plane optical (ZO) phonon mode, in such a way that the frequency of its overtone decreases to $2\omega_{\rm ZO}\sim\omega_{\rm G}$, 
where $\omega_{\rm G}$=1585~cm$^{-1}$ is the frequency of the Raman-active E$_{\rm 2g}$ mode.
Thus, the observed new peak is attributed to the 2ZO mode which becomes Raman-active following a mechanism known as Fermi resonance.
First-principles calculations on vibrational and anharmonic properties of the graphene/Cr interface support this scenario.

\end{abstract}
%%  78.30.Na Infar-red and Raman spectra: Fullerenes and related materials
%%  63.22.-m Phonons or vibrational states in low-dimensional structures and nanoscale materials
%%  71.15.Mb   Density functional theory, local density approximation, gradient and other corrections
\maketitle

\section {Introduction}

Raman spectroscopy of graphene and graphitic systems is extensively studied and, as example, it is commonly used to quantify disorder or to determine the number of layers in a graphene flake~\cite{ferrari04}.
In the case of graphene, Raman spectroscopy has proven highly sensitive to degree of disorder, mechanical strain and electronic doping~\cite{malard09, pimenta07, ferrari13, lee12, reserbat12, bronsgeest15}.
Such sensitivity originates from the strong and unusual electron-phonon coupling in the material~\cite{Pisana07,Yan07} and to the anharmonicity of the interatomic potentials~\cite{Bonini07}, which determine the lineshape and frequency of the Raman modes.
The contact of graphene with foreign species may have drastic consequences on its vibrational properties.
Strong interaction with a metallic surface for instance modifies electron-phonon coupling~\cite{allard10} in such a way that the Raman signal may be suppressed~\cite{Sutter08,Starodub11}, while charge transfers~\cite{Lee10,Wang11} and interfacial stress~\cite{usachov18} can substantially shift the frequency of the modes.
In spite of the huge amount of works already done, today these systems can still display unexpected phenomena as, \textit{e.g.} when graphene interacts with metallic systems~\cite{usachov18}.

Fermi resonance is a phenomenon observed in Raman (and infra-red) spectroscopy, first described in the '30s~\cite{fermi31}. It concerns an atomic vibration with frequency $\omega_1$, whose double excitation
(first overtone) would normally have a small Raman intensity. 
%Note: It does not really matter the Raman intensity of omega_1, we are talking about the overtone here
When the energy of the overtone matches the energy of a second Raman active vibration ($2\omega_1\simeq\omega_2$), the modes can mix and the Raman intensity of the overtone can become important. In Fermi's seminal work~\cite{fermi31}, the mixing originates from anharmonic coupling of strength $V$ between the vibrations, and it becomes significant when $V \sim \hbar|2\omega_1-\omega_2|$ (Ref.~\onlinecite{fermi31}).
So far, this phenomenon has been mainly observed in molecules and molecular solids (\textit{e.g.} CO$_2$ and organic compounds~\cite{agranovich}) and to our knowledge not in dispersive simple crystals.

In this work we reveal an anomalous Raman spectrum for graphene grown on a chromium-coated copper foil. We observe an intense peak with a frequency of 1608~cm$^{-1}$ (named as `U' peak for `unknown origin' peak), which has not been described before to our knowledge.
Strikingly, the new peak has an intensity comparable to that of the well-documented most-prominent features in the
Raman spectrum of graphene and its presence is indeed extraordinary, considering the
huge amount of Raman experiments published on graphene and of the different experimental
conditions exploited so far.
We find that the new peak is not dispersive and not linked to the presence of defect, differing in this sense from the close-in-energy D$'$ peak. Moreover, it does not originate from strain or electronic doping effect. Based on density functional theory (DFT) calculations, we argue that the peak is related to the overtone of the out-of-plane optical ZO mode of graphene, whose intensity is largely increased through a Fermi resonance mechanism
involving anharmonic mixing with the Raman active E$_{2g}$ phonon mode. Such mixing is made possible by the presence of chromium (Cr), which softens the ZO phonon modes in the relevant frequency range,
at variance with what expected from other previously studied materials such as copper or nickel.

The paper is organized as follows:
Sect.~\ref{sec1} describes the growth and structure of the samples; Sect.~\ref{sec2} is devoted to our Raman spectroscopy analysis; Sect.~\ref{sec3} describes the computational approach; finally
Sect.~\ref{sec4} discusses the Fermi resonance hypothesis.

\section{Growth and structure of the samples}
\label{sec1}

\subsection{Graphene growth on different substrates and transfer to SiO$_2$}
\label{Growth information}
\begin{figure*} [!bth] 
\resizebox{2.0\columnwidth}{!}
%\centering
%\includegraphics[width=1\textwidth]{FigS1}
{\includegraphics{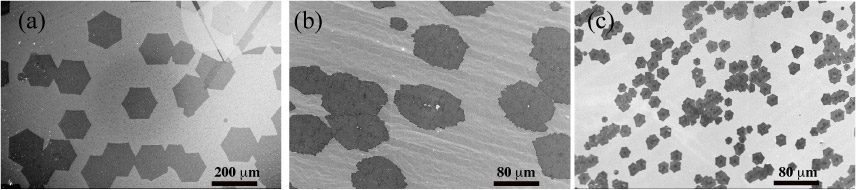}}
\caption[width=1\textwidth]{SEM images of graphene during the CVD growth on three different substrates:
a) Uncoated 99.999\% pure copper without annealing; b) Cr coated 99.8\% pure copper without annealing;
c) Cr coated 99.8\% pure copper with annealing in H$_2$ atmosphere at 1 mbar for 1 hr.}
\label{fig_SEM}
\resizebox{1.7\columnwidth}{!}
%\centering
%\includegraphics[width=1\textwidth]{FigS2.eps}
{\includegraphics{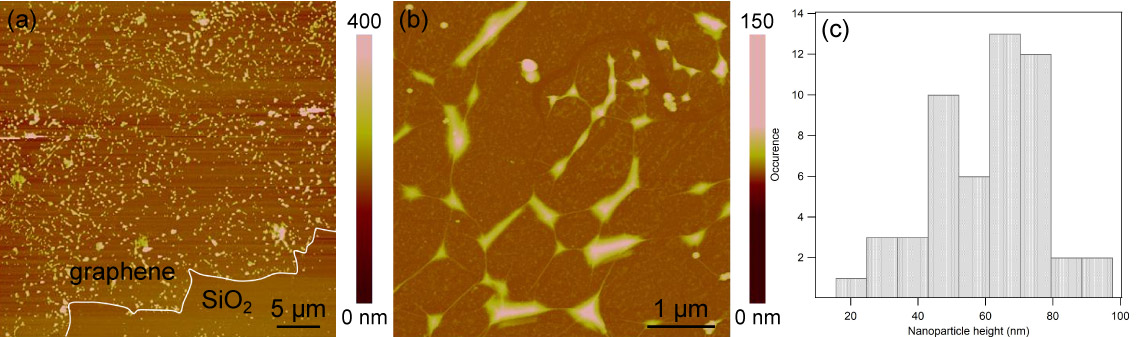}}
\caption{Atomic force microscopy (AFM) images of nanoparticles
underneath graphene, near a graphene edge (a), and near the center of
a graphene island (b). Smaller nanoparticles are prominently found in
the center of the islands. Micrometer-scale features coexist with the
nanoparticles. Wrinkles are associated with the presence of these
features, which indicates they indeed stand underneath graphene
(see discussion).
%Some torn parts of graphene (red arrow) are also visible; they are
%presumably induced by the transfer process. 
(c) Size distribution of the nanoparticles.}
\label{fig_AFMsizeDist}
\end{figure*}

Chemical vapour deposition (CVD) of graphene was performed inside a quartz
tube heated to 1050$^\circ$C. Methane was used as a carbon precursor,
as part of a gas mixture with hydrogen and argon (in 4:1000:500 sccm
proportions). Methane was let in for 15~min, immediately after
1050$^\circ$C was reached, presumably at a stage when the Cu foil
(99.999\% purity, supplier: Alpha Aesar) used as a substrate is still
covered with its oxide. In such growth conditions and on such substrates, graphene [referred to as of type (i) in the following] has been shown to form with a very low nucleation density, corresponding to holes in the copper oxide [Fig.~\ref{fig_SEM}(a)]. This leads to the slow growth of large,
far-apart, single-crystal graphene islands \cite{zhou13}. A typical
scanning electron micrograph of the Cu surface after growth and
cool-down is shown in Fig.~\ref{fig_SEM}(a). The average graphene island
size is about 170~$\mu$m.

Copper foils also exist that are protected from (copper) oxidation by
a passivating chromium-chromium oxide coating (here, Cu purity:
99.8\%, supplier: Alpha Aesar -- product no. 13382). It appears that
the Cr-based coating is also largely passivating the Cu
surface at 1050$^\circ$C against graphene formation, leaving, like in
the above-discussed case of copper oxide, few holes where graphene
nucleation can take place [Fig.~\ref{fig_SEM}(b)]. We expect this coating to
be essentially metal Cr, given the presence of hydrogen which
will quickly reduce the thin passivating oxide.
A Cr coating is expected to hinder graphene growth (no CVD of graphene on Cr was
reported to our knowledge) and acts as the copper oxide discussed in
the previous paragraph. Graphene growth hinderance appears less
efficient than with copper oxide, though, and we observe a slightly
higher nucleation density for graphene together with a smaller size of
the graphene islands, of 80~$\mu$m [Fig.~\ref{fig_SEM}(b)]. We also note
that the shape of the graphene islands is less symmetric, indicative
of a non-single-crystal nature. Graphene grown in these conditions and on such substrates are referred to as of type (ii) in the following.

A third kind of substrate was also used, once more a chromium-chromium
oxide-coated copper foil (Cu purity: 99.8\%, supplier: Alpha Aesar)
but this time annealed at 1050$^\circ$C for 1~h under 1~mbar of H$_2$. On such a
substrate we observe a much higher nucleation density for graphene
together with a much smaller island size, typically of the order of
10-20~$\mu$m. The islands have a highly symmetric shape, indicative of
their single-crystal nature [Fig.~\ref{fig_SEM}(c)]. These attributes are
typical of graphene grown on the clean surface of copper foils. We
argue that these observations indicate that, after rapid chromium
oxide reduction in presence of hydrogen, the resulting metal Cr
dissolves in the copper substrate [Fig.~\ref{fig_SEM}(c)] -- the 1~h
annealing leaving sufficient time for this slow process to occur. Graphene grown in these conditions and on such substrates are referred to as of type (iii) in the following.

After the growth, all samples are cooled down to room temperature in the
presence of 100 sccm of argon.
Graphene transfer from the Cu substrates to SiO$_2$/Si is performed
using polymethyl methacrylate (PMMA) as a thin film withholding
graphene, as described in Ref.~\onlinecite{han14}. The PMMA is
spin-coated on the sample and heated to 180~$^\circ$C for
1~min. Further, the Cu foil is etched using a 0.1~mg/mL solution of
(NH$_4$)$_{2}$SO$_4$. The floating graphene+PMMA is washed in
de-ionized water and fished on a 300 nm SiO$_2$/Si piece of wafer,
onto which it is left to dry. Finally the transferred sample is heated
at 120$^\circ$C for few minutes before dipping into an acetone bath
dissolving PMMA.

\subsection{\label{secNPs} Chromium-based nanoparticules in contact with graphene}

Graphene prepared on Cr-coated copper foils [without H$_2$ annealing,
Fig.~\ref{fig_SEM}(b)] comes together with high density of nanoscale features, as observed in atomic force microscopy (AFM, Fig.~\ref{fig_AFMsizeDist}). Strikingly, some of the nanoparticles appear elongated and seem to be prolonged at their ends by pleats, as the height gradually vanishes to the level of the graphene plane [Fig.~\ref{fig_AFMsizeDist}(b)]. We deduce that these features are actually nanoparticles standing underneath the graphene cover, acting much like masts under a graphene tent, suspending graphene and yielding pleats at the edges of a graphene portion suspended by \textit{e.g.} two nanoparticles.

These nanoparticles are transfered together with graphene onto the SiO$_2$/Si substrate. Atomic force microscopy reveals nanoparticles with average size of 56~nm, as shown in Fig.~\ref{fig_AFMsizeDist}(c). We find that the density of nanoparticles is higher closer to the edges of the graphene islands, increasing from 2.5~$\mu$m$^{-2}$ (center) to 6.2~$\mu$m$^{-2}$ (edge). This increased edge density points to an outward diffusion of the nanoparticles induced by graphene growth, suggesting that the graphene growth front drags the nanoparticles. The reason for this effect is unknown at the moment.

These particles are obviously linked to the presence of Cr at the copper surface. We note that the Cr coating~\cite{nota01} is initially (at room temperature) covered by a thin (typically few-nm-thick) Cr-oxide layer. The oxide is however readily reduced by hydrogen at the elevated temperatures~\cite{chu79} used in the graphene growth process and the nanoparticles are then likely to be metallic~\cite{nota02}.

\begin{figure} [!htb]
\begin{center}
\includegraphics[width=0.70\columnwidth]{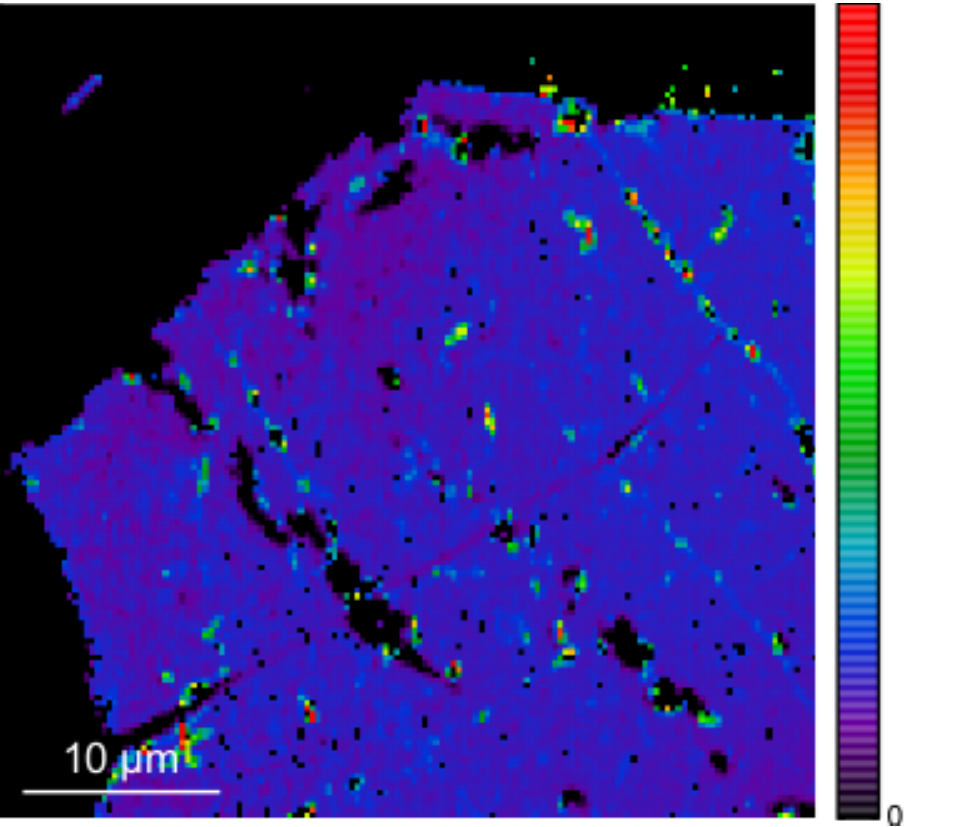}
\end{center}
\caption{Spatial dependency of the Raman G line area, I(G) [arbitrary units]. The zone is the same as the one
depicted in Fig.~\ref{fig2}(b).} 
\label{fig_gmap}
\end{figure}

Fig.~\ref{fig_gmap}, reports a characteristic spatial Raman imaging of the area
of the G peak, I(G), which turns out to be quite homogeneous all over the maps that were taken. Because of this, the maps with relative area I(U)/I(G) and I(D)/I(G) that will
appear on the discussion for the U and D peaks are not essentially different from the corresponding I(U) and I(D) maps.

%Black pixels correspond to areas where the fit of the G peak is impossible due to the presence of spikes.
%In the zone where there are nanoparticles, the area of the peak G is reduced by almost 30\%. Knowing that the area ratio I (2ZO) / I (G) is 5-6 in these areas, it contributes marginally to the magnitude of this factor.

%\nedjma{PROBABLY WE NEED TO ADD A COMMENT ABOUT CHEMICAL SIGNATURE OF CHROME. THE EDX SHOW THAT THERE IS CHROMIUM BUT WE DON'T HAVE ANY INFORMATION ABOUT ITS METALLICITY OR NOT.}

\section{Raman spectroscopy in the presence/absence of nanoparticles}
\label{sec2}

\subsection{Observation of new Raman peak}

\begin{figure}
\includegraphics{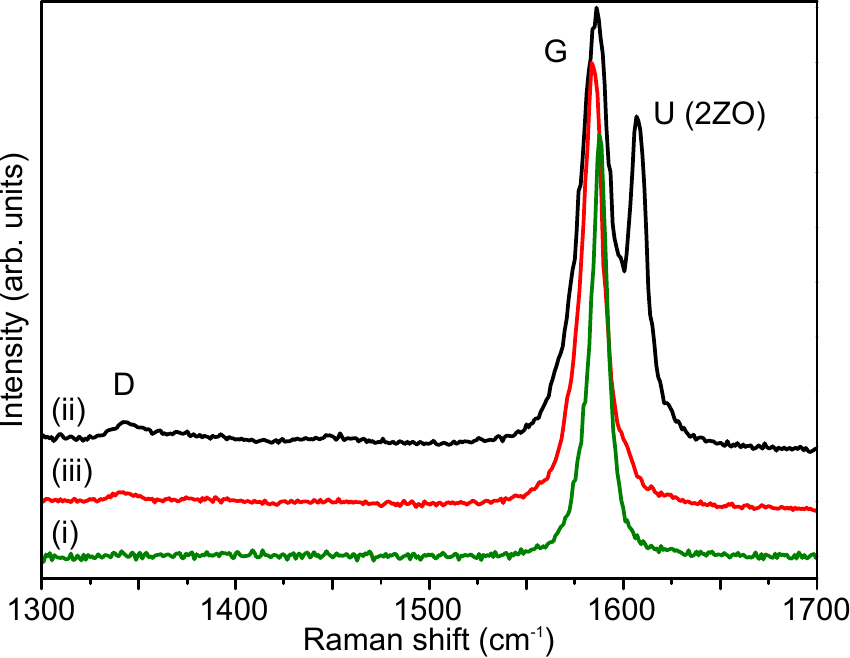}
\caption{(Color online)
Raman spectra of graphene grown on (i) a Cr-free Cu foil,  (ii) a Cr-coated non-annealed Cu foil, and (iii) an annealed Cr-coated Cu foil.
The spectra have been vertically shifted for clarity. The exciting laser energy is 2.33~eV ($\lambda=532$~nm).
}
\label{fig1}
\end{figure}

Figure~\ref{fig1} displays representative Raman spectra taken on a SiO$_2$ substrate for the three kinds of graphene samples we introduced in section~\ref{Growth information}, \textit{i.e.} grown on (i) a Cr-free Cu foil, (ii) a Cr-coated Cu foil, and (iii) the same foil annealed to 1050$^\circ$C for 1~h in a H$_2$ reducing atmosphere. 
For graphene samples of type (i) and (iii), the Raman spectra show no remarkable difference with those commonly reported in the literature~\cite{ferrari13}. The most prominent peak, the G mode, is centered at $\omega_\mathrm{G}$ = 1584 and 1588~cm$^{-1}$ for type (i) and (iii) samples respectively. Their full-width at half maximum (FHWM) are 13 and 10~cm$^{-1}$ respectively. 
For type (ii) samples, prepared with a Cr-coated non-annealed Cu foil, the Raman spectrum is strongly anomalous. Indeed, beside the usual G peak ($\omega_{\rm G}$=1585~cm$^{-1}$, FWHM $\sim $ 18~cm$^{-1}$) and a defect-activated D peak ($\sim$1350~cm$^{-1}$) with low intensity, the spectrum displays an intense peak at 1608~cm$^{-1}$ [referred to as U(2ZO) peak for reasons that will become clearer later, Fig.~\ref{fig1}], with FWHM $\sim $ 9~cm$^{-1}$.

The occurrence of the new peak is linked to the presence of the Cr nanoparticles that we analysed in section~\ref{secNPs}.

\begin{figure}
\includegraphics{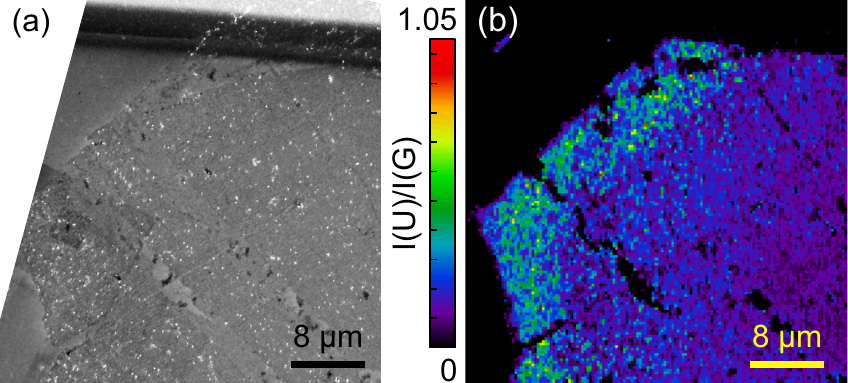}
\caption{(Color online)
(a) Scanning electron micrograph of graphene prepared on non-annealed Cr-coated Cu, and then transferred to SiO$_2$/Si. Graphene is essentially single-layer, except for darker regions which are multilayer patches. (b) Spatial Raman imaging of the area ratio of the U (2ZO) and G peaks.}
\label{fig2}
\end{figure}

Another key point is that the density of nanoparticles is substantially higher near the edge of the graphene flake (6.2 per $\mu$m$^2$) than at its center (2.5 per $\mu$m$^2$). This increased density is correlated with an increased intensity towards the flake edge of the new peak as measured by spatially localized Raman [Fig.~\ref{fig2}(b)]. This spatial correlation further confirms the central role of the nanoparticles in the occurrence of the new peak.

In the remaining of the current section, we will consider several possible origins for the new peak. For that purpose we turn to a spatially-resolved analysis of the correlations between the occurence/intensity of the new peak on one hand, and the presence of mechanical strain, charge carrier density, and point defects on the other hand. This analysis relies on hyperspectral Raman imaging, that we present in different flavours.

\subsection {Strain and charge carrier concentration imaging}

\begin{figure}
\includegraphics{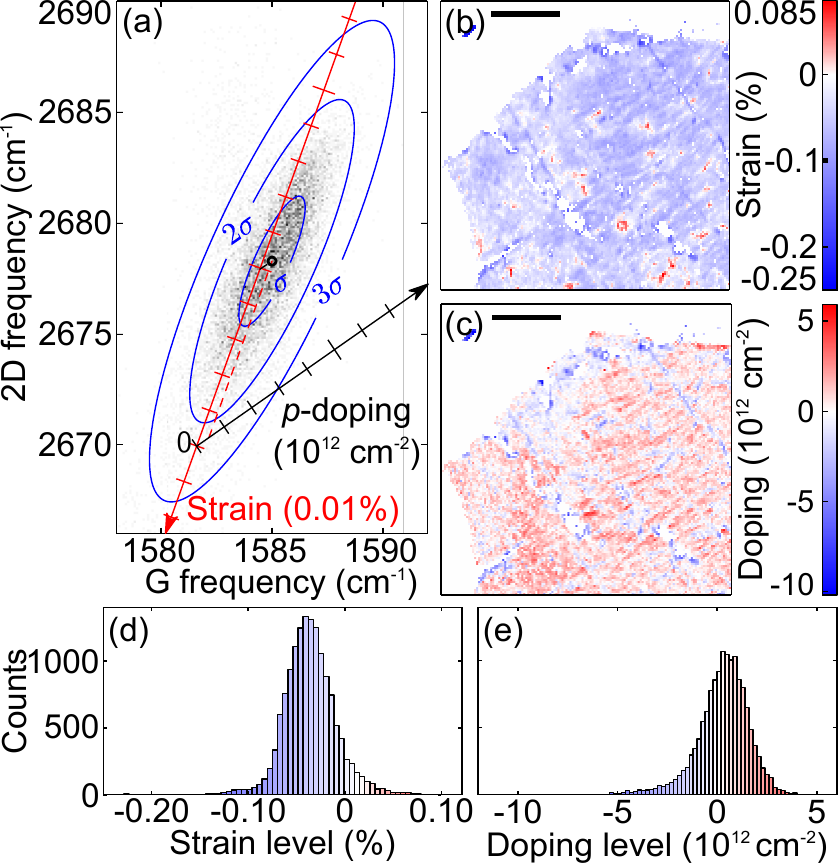}
\caption{(Color online)
(a) Frequency of the Raman 2D and G peaks measured on the same graphene flake analyzed on Fig.~\ref{fig2}. The ellipses delimit the 1$\times\sigma$, 2$\times\sigma$, and 3$\times\sigma$ ($\sigma$: standard deviation) widths of a Gaussian distribution fitted to the data and centered around the 1585 and 2678~cm$^{-1}$ average frequencies of the G and 2D modes.
The G and 2D frequencies corresponding to zero doping and strain~\cite{lee12} are 1581 and 2670 cm$^{-1}$ (the laser energy is 2.33~eV). (b,c) Map of the local strain and doping of the flake, obtained by applying the approach of Ref.~\cite{lee12} to the data of panel (a). (d,e) Histogram representation of the data from panels (b) and (c).}
\label{fig3}
\end{figure}

We first measure the spatial variation of the G and 2D modes frequencies, Fig.~\ref{fig3}(a). These frequencies each depend quantitatively on strain and doping in a distinctive, well-characterized manner~\cite{Pisana07, Yan07, Mohiuddin09}. The specific dependencies have been confirmed for graphene samples on different substrates\cite{Bendiab18} and with different defect concentrations. Representing each location on a graphene sample by a point with the G and 2D peak frequencies as coordinates hence allows to deduce, location-by-location, the local strain and charge carrier concentration~\cite{lee12}, and in turn, to construct a spatial map of these two quantities. 
The 2D frequency \textit{versus} G frequency scatter plot shown in Fig.~\ref{fig3} reveals a spatially-varying strain, that we relate to a compressive strain of the order of 0.1\%. These variations are visualized on Fig.~\ref{fig3} (b).

The charge carrier density also shows spatial variations [Fig.~\ref{fig3}(c)] ($\sim$1$\times$10$^{12}$ per cm$^2$ on average), that may also be related to the presence of the Cr nanoparticles, or to the more or less intimate contact with the SiO$_2$ surface, or to the presence of randomly distributed residues from the transfer process (from the growth substrate to SiO$_2$).
The strain and charge carrier concentration maps show essentially no correlations Fig.~\ref{fig3}(b,c). 
As already mentioned, the intensity of the new peak correlates with the concentration of nanoparticles [Fig.~\ref{fig2}(b)], which is higher toward the edges of the flakes, but shows no particular correlation with the strain and charge carrier density maps. Overall, these are usual values of strain and doping and, it is clear that we are not in the presence of an anomalous situation and that the new peak can not be attributed to these effects.

%%%%%Added on resubmission (3/6):%%%%%%%%%%%%%%%%%%%%%%%%%%%%
We additionally note that, when it is observed, the new peak always appears as a well-defined peak, distinct from the G mode, displaying an overall uniform frequency.
It is therefore unlikely that it results from locally large mechanical strains (not detected in the analysis discussed above) that would split the G peak.

Moreover, pseudomagnetic fields ranging from tens to hundreds of Teslas are expected in graphene deformed by nanoparticles underneath it~\cite{yamamoto12}. Such pseudo-magnetic fields would lead to an additional component of the 2D peak~\cite{faugeras10}, which we do not observe.
%%%%%%%%%%%%%%%%%%%%%%%%%%%%%%%%%%%%%%%%%%%%%%%%%%

\subsection {Role of defects in the Raman processes}

\begin{figure}
\includegraphics{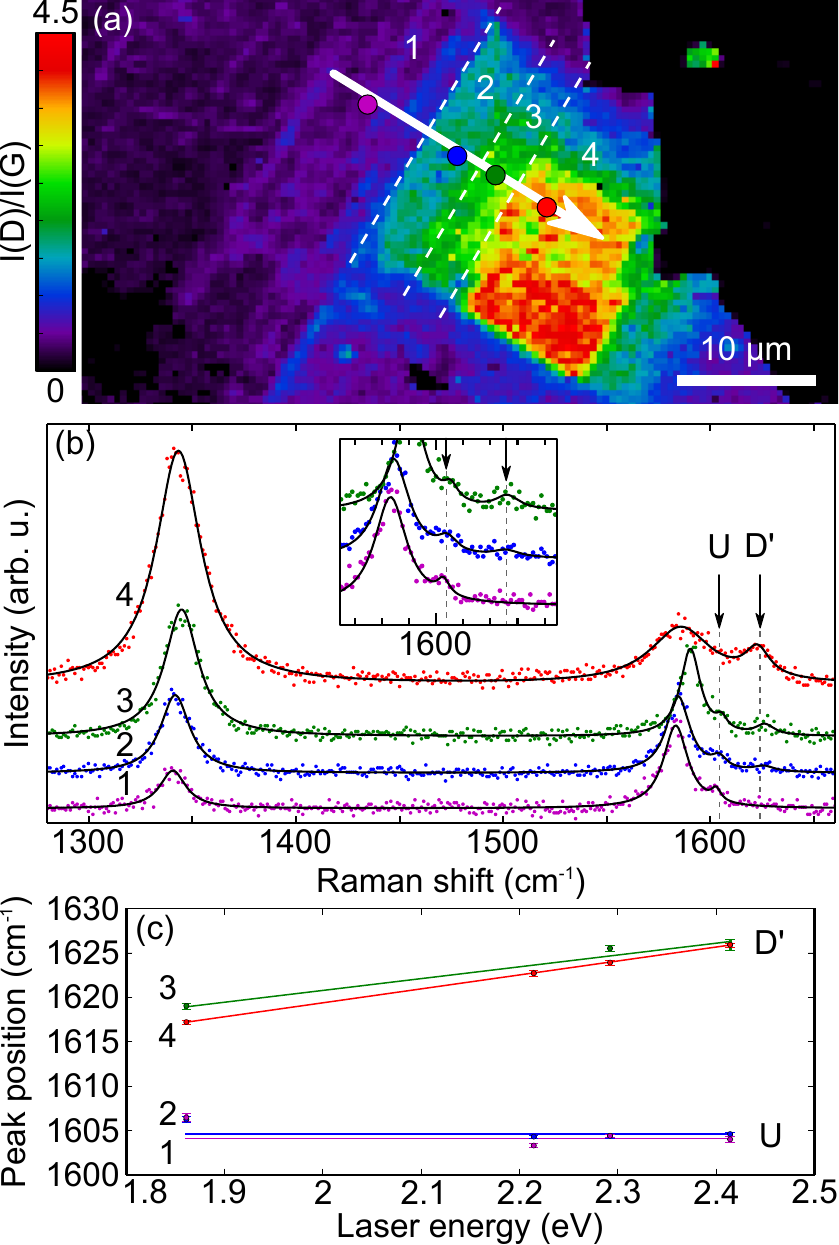}
\caption{(Color online) D$'$ and the new peaks in graphene. (a) Area ratio of the D and G peaks mapped over an area of the sample with zones of increasing electron beam irradiation. (b) Raman spectra acquired at different locations, exposed to an increased electron dose (numbered from 1 to 4). Inset: zoomed-in view of the D$'$ peak and the new peak (U). (c) Frequency of the D$'$ and U peaks as a function of excitation laser energy. The analysis is done at locations 1 and 2 in (a) for the new peak, and at locations 3 and 4 in (a) for the D$'$ peak. The result of linear fits to the D$'$ and new peak frequencies \textit{versus} laser energy are overlaid on the data points. The horizontal lines are guides to the eye.}
\label{fig4}
\end{figure}

The new peak's frequency falls close to the value expected for the defect-activated D$'$ peak, usually reported close to 1620~cm$^{-1}$~$^{1-4}$. The D$'$ peak is commonly attributed to the excitation of a small wavevector phonon of the LO branch through the so called ``double resonance'' mechanism. It has always been observed in the presence of a strong intensity D peak at 1350~cm$^{-1}$ ~$^{1-4}$, while our observations only reveal a weak D peak as evidenced in Fig.~\ref{fig1}. To definitely exclude the assignment of the new peak as the D$'$ peak, we intentionally generated defects into the sample.
For that purpose we bombarded the graphene sample with the electron beam of a scanning electron microscope (SEM), and increased step-by-step the area of the sample scanned by the beam to create regions exposed to a gradually increasing electron dose, hence exhibiting an increasing defect density~\cite{Teweldebrhan09} [Fig.~\ref{fig4}(a)].

We observe an increase of the D peak intensity with the defect density, and further the emergence of the D$'$ peak at 1620~cm$^{-1}$ [data set numbered 1$\rightarrow$4 in Fig.~\ref{fig4}(a,b)].
The D$'$ peak and the new peak are well separated. Besides, while the D$'$ frequency displays as expected~\cite{pimenta07} a marked dependence on the energy of the excitation laser (with a shift of 14$\pm$1~cm$^{-1}$ per eV on the explored range), the new peak frequency remains essentially constant.

The new peak and the well-documented D$'$ peak can thus be considered as signatures of different phenomena.

\section{Computational approach}
\label{sec3}

\subsection {Structure and frequency calculations}

\begin{figure}
\includegraphics[width=7.0cm]{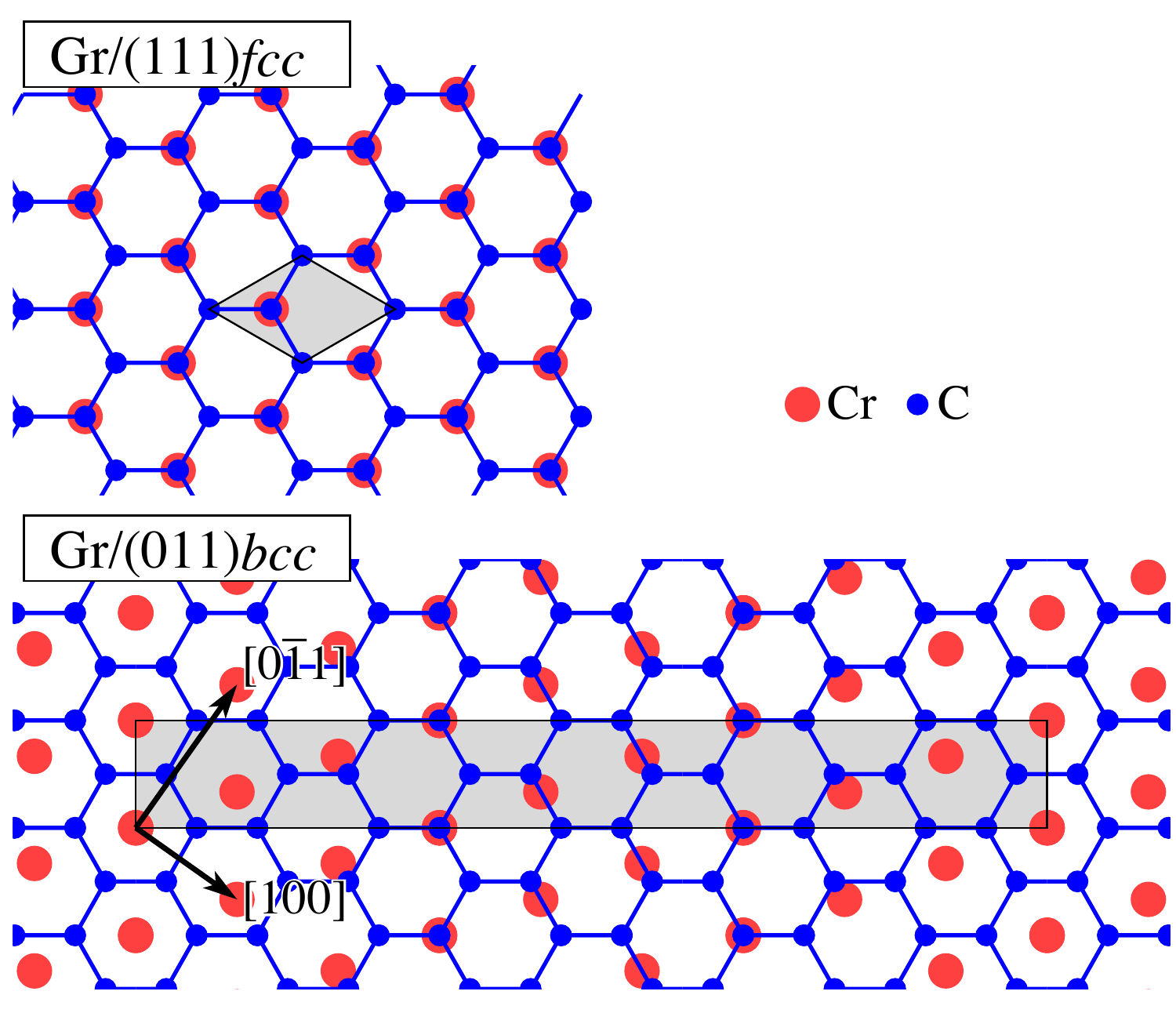}
\caption{
(Color online) Scheme of the graphene/Cr interfaces (top view) used for calculations.
Graphene (Gr) is lying on the (111) surface of $fcc$ Cr (upper panel) or
on the (011) surface of $bcc$ Cr (lower panel).
Only the Cr atoms of the outermost surface layer are shown.
Shaded areas: unit cell.}
\label{fig_structure}
\end{figure}

We calculated the phonon frequency of the zero wavevector ZO phonon mode 
of graphene,  $\omega_{\rm ZO}$, when graphene is in contact with three different metals (Cr, Cu, and Ni).
Calculations were done within density functional theory (DFT)  by using
the quantum-espresso package~\cite{giannozzi09} (details are in Appendix ~\ref{appendix1}).
By using periodic boundary conditions, three kinds of structures were simulated,
with graphene lying on
i) the (111) surface of a bulk $fcc$ metal
(the in-plane lattice spacing of graphene and that of the surface are matching, Fig.~\ref{fig_structure});
ii) one single layer of the $fcc$ (111) surface 
(\textit{i.e.}, the atoms of the metal are in the triangular close packed arrangement);
iii) one single layer of the $bcc$ (011) surface
within the reconstruction depicted in Fig.~\ref{fig_structure} (20 C atoms and 9 Cr atoms per cell).
Results are in table~\ref{tab_calcs}.

Cu and Ni were chosen as reference systems that are well characterized experimentally and theoretically
(see Refs.~\onlinecite{xu10,khomyakov09,allard10} and Refs. therein).
Note that, in spite of the strong graphene/Cr bond, with an
equilibrium distance very similar to that of Ni (table~\ref{tab_calcs}), the two associated
$\omega_{\rm ZO}$ are remarkably different.
Note also that $\omega_{\rm ZO}$ obtained for graphene lying on Cr $bcc$ and $fcc$ surfaces
are relatively similar (Table ~\ref{tab_calcs}).
This happens in spite of the differences between the atomic arrangement of the $bcc$ and $fcc$ surfaces.
Besides the fact that the (011) $bcc$ surface is slightly more open than
the (111)$fcc$, the relative alignment of the C and Cr atoms is substantially different:
while for the Gr/$fcc$ interface, half of the C atoms are on the top of a Cr atom,
in the Gr/$bcc$ one only one C out of 10 is aligned on the top of a Cr (Fig.~\ref{fig_structure}).

\begin{table}
\begin{tabular}{l |c|c||c|c|c||c|c}
   & \multicolumn{2}{c||}{Cu$^{fcc}$}& Cr$^{bcc}$ & \multicolumn{2}{c||}{Cr$^{fcc}$} & \multicolumn{2}{c}{Ni$^{fcc}$} \\
   & blk & 1L & 1L & blk &1L & blk & 1L \\
   \hline
   $\omega_{\rm ZO}$ (cm$^{-1}$)         & 878      & 869    & 767     & 771    & 788    & 716    & 688    \\
 $d$ (\AA)                                          & 3.23     & 3.00   & 1.95    & 1.99   & 1.92   & 2.06   & 2.05   \\
 $\Delta d$ (\AA)                               & 0.0005 & 0.001 & 0.035  & 0.043 & 0.045 & 0.003 & 0.009
\end{tabular}
 \caption{
Calculated $\omega_{\rm ZO}$ for graphene lying on three different metals:
bulk (blk) or only one layer (1L), with $fcc$ or $bcc$ structure.
$d$ is the average equilibrium distance of the C atoms from the metal plane. 
$\Delta d$ is the square root of the mean square vertical displacement of the C atoms from
the graphene plane.}
\label{tab_calcs}
\end{table}

\subsection {Anharmonic coefficients calculation}
\label{sec_anharmonic}

We determined within DFT the anharmonic three-phonon scattering
coefficients among the graphene phonon modes E$_{2g}$ ({\bf q=0}) and
two ZO phonons with wavevectors {\bf q} and {\bf -q}
and frequency $\omega_{\bf q}^{\rm ZO}$:
\begin{equation}
V_\alpha({\bf q})=\frac{1}{2}\sqrt{\frac{\hbar}{m\omega_{\rm G}}}
\left.\frac{\partial \hbar\omega^{\rm ZO}_{\bf q}}{\partial l}\right)_\alpha,
\label{eq_V0perfect}
\end{equation}
where $m$ is the carbon mass, $\omega_{\rm G}$=1585~cm$^{-1}$ 
and the derivative of $\omega_{\rm ZO}$ is done with respect to an atomic
displacement following the E$_{2g}$ phonon pattern in which 
each atom is displaced by the length $l$.
The index $\alpha=1,2$ selects the polarization of the doubly degenerate E$_{2g}$ mode.
Comparing with the notation of Eq.~4 of Ref.~\onlinecite{paulatto13},
$V_\alpha({\bf q})$ is equal to $V^{(3)}_{{\bf 0}j, {\bf q}j',{\bf -q}j'}$, where $j$ corresponds to
the E$_{2g}$ mode $\alpha$, and $j'$ to the index of the ZO mode.

The anharmonic interaction in graphene has been extensively studied (see \textit{e.g.} 
Ref.~\onlinecite{bonini07}) and for free standing graphene we calculated
$V_\alpha({\bf q})$  coefficients with Ref.~\onlinecite{paulatto13}'s approach.
For the two graphene/Cr structures depicted in Fig.~\ref{fig_structure}, we directly used
Eq.~\ref{eq_V0perfect} and finite differentiation
(each atom was displaced by 0, $\pm$0.02~\AA, $\pm$0.05~\AA ~and the five $\omega_{\rm ZO}$ thus
obtained were fitted with a polynomial to obtain the derivative).
$V_\alpha({\bf q})$ is an energy and to simplify the comparison with the Raman data will be expressed 
as a wavenumber in cm$^{-1}$.
For the E$_{2g}$ mode polarized along the vertical axis of Fig.~\ref{fig_structure},
$V_\alpha({\bf 0})$=0.0 for the Gr/$fcc$ structure because of the horizontal mirror symmetry.
For the E$_{2g}$ mode polarized along the horizontal axis of Fig.~\ref{fig_structure}
$V_\alpha({\bf 0})$=0.5 cm$^{-1}$ and $V_\alpha({\bf 0})$=6.0 cm$^{-1}$ for the
Gr/$fcc$ and Gr/$bcc$ structures, respectively.

\subsection {Model for the Fermi resonance}
\label{sec_Fermi_model}

Here, we describe a model to study, in graphene, the anharmonic mixing
among the doubly degenerate {\bf q=0} E$_{2g}$ phonon mode and
two ZO modes with wavevector {\bf q} and {\bf -q}.
The angular frequencies will be indicated as $\omega_{\rm G}$ for the E$_{2g}$
and $\omega_{\bf q}$ for the ZO and we will suppose to be in the presence of an external
mechanism that shifts the phonon frequencies so that $\omega_{\rm G}\simeq 2\omega_{\bf 0}$.
The generic Hamiltonian describing the anharmonic phonon-phonon interaction for an extended system has the
form reported, e.g. in Ref.~\onlinecite{menendez84}.
The Hamiltonian, restricted to the phonons of interest, is
\begin{eqnarray}
{\cal H}&=&
\sum_\alpha \left( a^\dagger_\alpha a_\alpha +\frac{1}{2}\right)\hbar\omega_{\rm G} + \nonumber \\
&&+\sum^*_{\bf q} \left( b^\dagger_{{\bf q}} b_{{\bf q}} + b^\dagger_{{\bf -q}} b_{{\bf -q}} + 1
\right)\hbar\omega_{{\bf q}}  + \nonumber \\
&&+\sqrt{\frac{2}{N'}}
\sum^*_{\alpha,{\bf q}} 
V_\alpha({\bf q}) A_\alpha B_{{\bf q}} B_{{\bf -q}},\label{eq1}
\end{eqnarray}
where $a^\dagger_\alpha$($a_\alpha$) is the creation (annihilation) operator associated
to the two E$_{2g}$ modes ($\alpha$=1,2) and
$b^\dagger_{\bf q}$ ($b_{\bf q}$) the creation (annihilation) operator
associated with a ZO phonon with wavevector {\bf q}.
$V_\alpha({\bf q})$ is defined in the previous subsection,
$A_\alpha=a^\dagger_\alpha+a_\alpha$ and $B_{\bf q}=b_{\bf q}+b^\dagger_{-q}$.
The sum over $\alpha$ is performed over $\alpha=1,2$ and
$\sum^*_{\bf q}$  is performed on a set of $N'$ wavevectors uniformly distributed within half
the two dimensional graphene Brillouin Zone (BZ).

To parametrize ${\cal H}$ we calculated  the phonon dispersion of the ZO mode
of free-standing graphene within DFT~\cite{giannozzi09}.
The results can be fitted in a large part of the BZ by $\omega_{\rm q} = -630 q^2 -1034 q^4$,
where we have shifted the energies so that $\omega_{\bf 0}$=0, $q=|{\bf q}|$,
$\omega$ is in cm$^{-1}$ and $q$ is in units of $2\pi/a_0$, being $a_0$ the graphene lattice spacing.
We chose $\omega_{\rm G}=-20$ ~cm$^{-1}$, which roughly reproduces the measured difference between the
G and U Raman peaks (see Sec.~\ref{sec4}).
For free standing graphene,  $V_\alpha({\bf q})$ can be approximated as
\begin{equation}
V_\alpha({\bf q}) = q^2 \left[ {\rm cos}(2\theta)\delta_{\alpha,1} +{\rm sin}(2\theta)\delta_{\alpha,2}  \right]
\times 36~{\rm cm}^{-1}, \label{eq_p1}
\end{equation}
where the two-dimensional vector {\bf q} is expressed as a function of the modulus $q$ and of the
direction defined by the polar angle $\theta$. $q$ is in units of $2\pi/a_0$.
Eq.~(\ref{eq_p1}) fits very well  DFT calculations for $q$ up to 0.5 and the specific dependence on $\alpha$
depends on the choice of the E$_{2g}$ mode polarization.
As an alternative we will also use:
\begin{equation}
V_\alpha({\bf q}) = \delta_{\alpha,1}   e^{-(q/\sigma)^2}  \times30~{\rm cm}^{-1} \label{eq_p2},
\end{equation}
whose meaning will be clarified in Sec.~\ref{sec4}.
Note that we are assuming that the anharmonic coupling is stronger in a specific direction 
for the E$_{2g}$ polarization (selected by $\alpha$=1) and, thus,
only one of the two E$_{2g}$ modes  is subject to the Fermi resonance,
while the second one ($\alpha$=2) is not.
%The coupling with the 2ZO mode will induce then a shift of the first E$_{2g}$ mode but not on the
%second, resulting in a splitting of the E$_{2g}$ mode, which will possibly appear as a broadening
%of the Raman G line.

Let us now consider the space ${\cal V}$ generated by
$|G_\alpha\rangle=a^\dagger_\alpha|0\rangle$ and
$b^\dagger_{{\bf q}}b^\dagger_{{\bf -q}}|0\rangle$,
which represents the excitation of two ZO phonons with opposite momentum.
To study how the $|G_\alpha\rangle$, which are Raman active,
are modified by the three-phonon interaction of Eq.~\ref{eq1} we calculated
the projected density of states
\begin{equation}
\chi_{G_\alpha}(\omega)=\sum_m |\langle G_\alpha|\psi_m\rangle|^2\delta(\omega-\omega_m),
\label{eq_chi}
\end{equation}
where $|\psi_m\rangle$/$\omega_m$ are the eigenstates/eigenvalues of ${\cal H}$.
$\chi_{G}(\omega)$ can be easily obtained with a Lanczos approach (see Appendix~\ref{appendix2}).

\section{Fermi resonance hypothesis}
\label{sec4}

The experimental results reported in Sec.~\ref{sec1} and~\ref{sec2}
can be used to discard several possible interpretation
for the new peak, but they do not provide indications on its origin, apart from the fact that
the contact with Cr nanoparticles seems a necessary condition for its emergence.
Here, we interpret the new U peak ($\omega_{\rm U}\simeq$~1608~cm$^{-1}$) as the overtone of the graphene ZO phonon mode (2ZO), arguing that the presence of Cr downshifts $\omega_{\rm ZO}$ from 890~cm$^{-1}$ for isolated graphene to 804 cm$^{-1}$. The frequency of the overtone (which is then $\omega_{\rm 2ZO}\simeq2\omega_{\rm ZO}\simeq$~1608~cm$^{-1}$) thus approaches the value of 1585~cm$^{-1}$ of the G peak corresponding to the E$_{2g}$ Raman-active mode. Because of the small frequency difference, the 2ZO and E$_{\rm 2g}$ modes can mix and the 2ZO becomes Raman active through the Fermi resonance mechanism$^{17,18}$.

Indeed, it is known that when graphene is in contact with a metal, $\omega_{\rm ZO}$ can downshift by hundreds of cm$^{-1}$ (see the measurements reported  in Fig.~\ref{fig_calcs}). 
This is not surprising since the ZO mode corresponds to an atomic vibration perpendicular to the graphene plane and even a partial hybridization of the graphene $p_z$ orbitals with the $d$ surface orbitals of the metal affects the related force constants~\cite{allard10, politano16}.
For certain metals, the value of $\omega_{\rm ZO}$ remains similar to that of isolated graphene, while for others $\omega_{\rm ZO}$ severely downshifts.
For example, according to the measurements of Ref.~37, $\omega_{\rm ZO}\simeq$~845~cm$^{-1}$ in the case of Cu, while $\omega_{\rm ZO}\simeq$~710~cm$^{-1}$ for Ni. One should note that graphene/Cu and graphene/Ni are usually classified as weakly and strongly bonded systems, respectively$^{23}$.
In these two cases and for the other occurrences reported in the literature~\cite{altaleb16} (Fig.~\ref{fig_calcs}) the frequency of the corresponding overtone 2$\omega_{\rm ZO}$ is too high or too low with respect to $\omega_{\rm G}$, and the Fermi resonance cannot be activated. This might explain why this phenomenon has not been observed before in spite of the many studies on graphene in contact with different substrates.

\subsection {ZO frequency shift from DFT}

The Fermi resonance hypothesis requires $\omega_{\rm ZO}\simeq$ 804~cm$^{-1}$ (half of the measured U peak frequency) which therefore implies that the graphene/Cr system displays an intermediate behavior between systems such as graphene/Cu and graphene/Ni.
To check this, we simulated the structure and vibrational properties of graphene/Cr using DFT ~\cite{giannozzi09}.
The precise atomic arrangement at the interface between the Cr nanoparticles and graphene is unknown and \textit{a priori} complex, and describing it accurately would require number of atoms that is not tractable with DFT calculations. We hence turn to a simplified description of the interface, assuming a flat graphene sheet lying on a flat well-defined Cr surface.

So far, relevant experimental and theoretical studies have focused on graphene/metal systems with dense-packed surfaces of hexagonal compact ($hcp$) and face-centered cubic ($fcc$) metals with lattice spacing comparable with that of graphene (see \textit{e.g.} Ref.~23), so that strain effects and reconstructions could reasonably be disregarded. On the contrary, Cr is stable in the body-centered cubic ($bcc$) structure. Its most compact surface is the (011), which is not commensurate with graphene. For the graphene/Cr(011) interface, the smallest reconstruction presenting a relatively small lattice mismatch ($<3\%$) is schematized in Fig.~\ref{fig_structure}.

\begin{figure}
\includegraphics{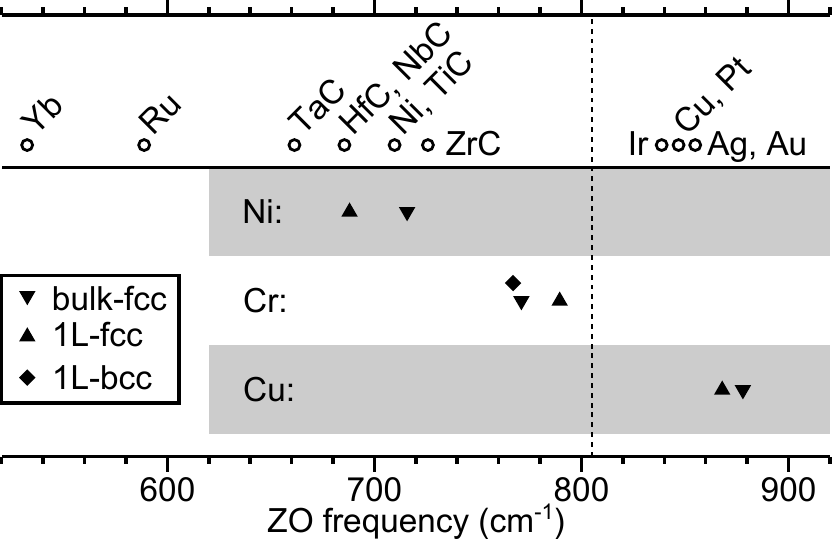}
\caption{$\omega_{\rm ZO}$ for graphene in contact with different materials. Open dots are measurements as reported in the review of Ref.~\onlinecite{altaleb16} (Cu and Ni experiments are from Ref.~\onlinecite{shikin98}).
Filled symbols represent present calculations done for graphene lying on different metals, which can be bulk crystals (bulk) or a single atomic layer (1L) with $fcc$ of $bcc$ structures.
The vertical dotted line corresponds to 804 cm$^{-1}$, which is half of the measured frequency of the U peak.
}
\label{fig_calcs}
\end{figure}

As a tentative approximation, we first consider this reconstruction with only one layer of Cr finding $\omega_{\rm ZO}=$788~cm$^{-1}$ 
This result is very encouraging, since 788~cm$^{-1}$ is not far from the value needed to activate the Fermi resonance.
To test the approach reliability, we also calculated $\omega_{\rm ZO}$ for graphene lying on the (111) surface of Ni, Cu, and Cr $fcc$ bulk metals and for graphene lying on one isolated layer of the three $fcc$ metals ($fcc$ Cr is metastable, with a lattice spacing compatible with that of graphene~$^{39}$). Results are reported in Fig.~\ref{fig_calcs}. The values for Ni and Cu bulk can be compared with measurements and the reasonable agreement confirms that DFT is predictive in this context. Moreover, the similarity between calculations done on the top of the bulk and on the top of an isolated layer suggests that,
for each of the three metals, $\omega_{\rm ZO}$ is mainly determined by the interaction with the outermost metallic layer. On the other hand, because of the similarity between the $\omega_{\rm ZO}$ frequency on the three Cr calculations (Fig.~\ref{fig_calcs}) we can argue that the actual atomic arrangement at the graphene/Cr interface plays a secondary role. Therefore, the present $\omega_{\rm ZO}$ calculation is at least qualitatively correct, and predicts that when graphene is in contact with Cr, $\omega_{\rm ZO}$ reaches a value compatible with the Fermi resonance scenario.

\subsection{Anharmonic mixing between the E$_{2g}$ and ZO phonons}
\label{Anharm}

Following the original idea of E. Fermi~\cite{fermi31}, 
if we assume that $2\omega_{\rm ZO}\sim\omega_{\rm G}$ (the frequency of the Raman active E$_{2g}$ mode)
the three-phonon anharmonic scattering term between two ZO and one E$_{2g}$ phonons could couple the modes
and activate the 2ZO overtone.
In isolated crystalline graphene, this term is zero for three {\bf q=0} phonons because of symmetry.
As a consequence, Raman activation can be possible only in the presence of a sufficiently strong degree of disorder.

To quantify the characteristics of this hypothetical disorder mechanism we numerically solved
the Fermi resonance model from Sec.~\ref{sec_Fermi_model}.
In particular, we considered two different parameterizations for the anharmonic interaction from
Eqs.~(\ref{eq_p1}) and (\ref{eq_p2}).
Eq.~(\ref{eq_p1}) corresponds to a perfectly crystalline graphene.
Eq.~(\ref{eq_p2}) is used to determine whether it exists a form for $V({\bf q})$
that can activate the Fermi resonance. 
In Eq.~(\ref{eq_p2}), $V({\bf 0})$ is different from zero and we have also introduced the parameter
$\sigma$ to limit the range in the reciprocal space of the {\bf q=0} component of the anharmonic coupling
(higher values of $\sigma$ result in a higher mixing between E$_{2g}$ and ZO).

\begin{figure}
\includegraphics[width=7.0cm]{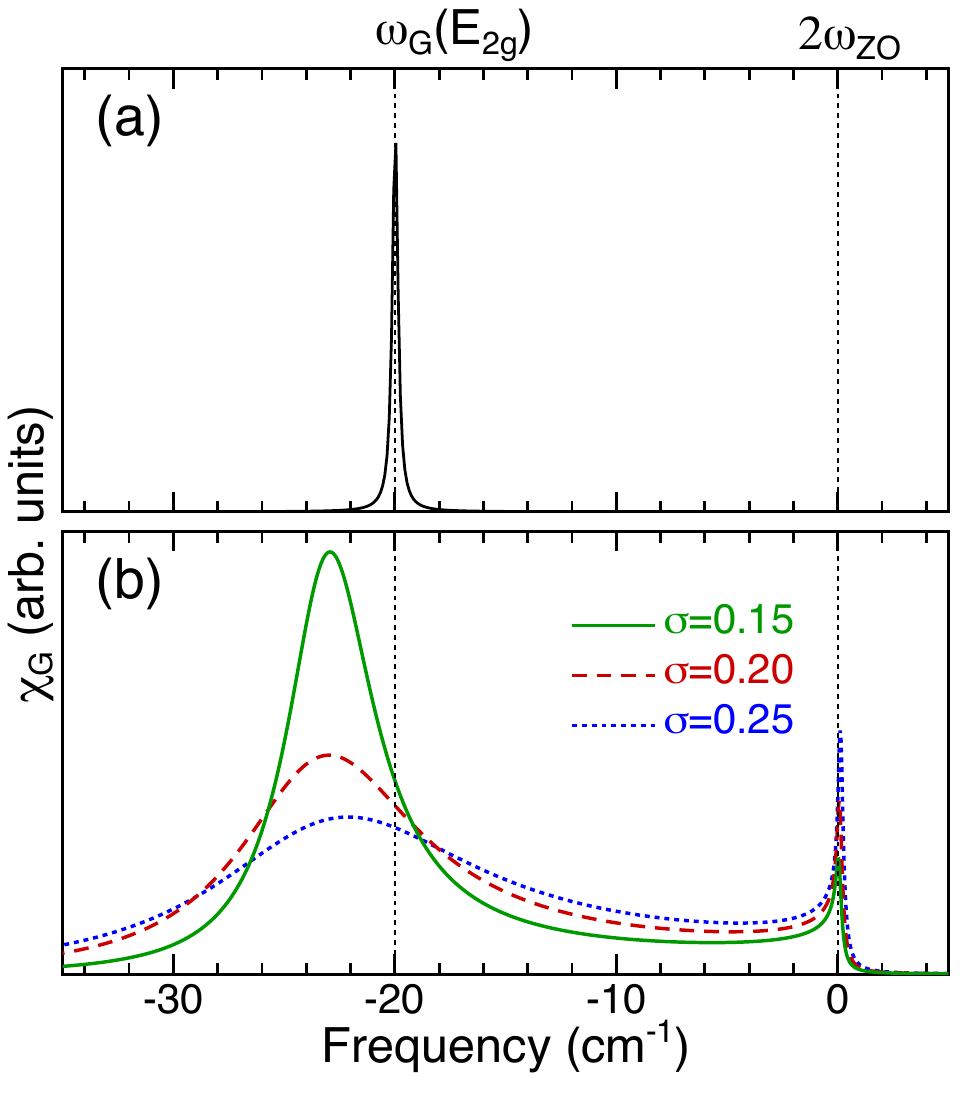}
\caption{
Calculated projected phonon density of states for the two models described by Eq.
~(\ref{eq_p1}) [panel (a)] and Eq.~(\ref{eq_p2}) [panel (b)] for different values of $\sigma$.
The vertical dotted lines correspond to the unperturbed frequencies
of the E$_{2g}$ mode, $\omega_{\rm G}$, and of the first overtone of the ZO mode $2\omega_{\rm ZO}$
(shifted to 0).}
\label{fig_chi}
\end{figure}

Fig.~\ref{fig_chi} displays the projected phonon density of states, $\chi_G$, Eq.~(\ref{eq_chi}),
obtained with the two alternative forms for the phonon-phonon interaction.
By using, Eq.~(\ref{eq_p1}), which corresponds to purely crystalline graphene, $\chi_G$
presents a single peak centered on the energy of the E$_{2g}$ mode ($\omega_{\rm G}$), meaning that the nature of the E$_{2g}$ mode is not substantially changed (apart from anharmonic broadening).
On the other hand, by using Eq.~\ref{eq_p2}, $\chi_G$ presents a second peak at the energy of the 2ZO
mode, meaning that the anharmonic interaction is sufficiently strong to couple the
2ZO with the E$_{2g}$ mode. The activation of the Fermi resonance is thus possible by invoking 
 the presence of a certain degree of disorder disrupting the intrinsic symmetries of graphene to such an extent that the E$_{2g}$/2ZO anharmonic coupling at {\bf q=0}, V({\bf 0}), is 30~cm$^{-1}$ [that is the one used in Eq.~(\ref{eq_p2}) and Fig.~\ref{fig_chi}(b), see Sec.~\ref{sec_Fermi_model}].

We argue that the contact to Cr could already be enough to induce such a coupling.
We first consider graphene lying on one layer of Cr $fcc$ atoms (Fig.~\ref{fig_structure}).
From DFT calculations (Sect. ~\ref{sec_anharmonic}), V({\bf 0}) = 0.5 cm$^{-1}$.
This is already a very encouraging result, meaning that 
the hybridation of C-Cr orbitals (which does not affect in a major way the atomic structure of graphene) 
is enough to break the symmetry to such a degree that the
anharmonic coupling is activated [otherwise V({\bf 0}) would remain zero].
The value 0.5 cm$^{-1}$ is relatively small with respect to 30 cm$^{-1}$,
however 0.5 cm$^{-1}$ is obtained for a highly symmetric structure in which 
one of the two C atoms of the graphene unit cell is exactly on the top of a Cr atom.
Indeed, if we consider graphene lying on one layer of Cr $bcc$ atoms  (Fig.~\ref{fig_structure}),
V({\bf 0}) becomes 6 cm$^{-1}$,
which is much closer to 30 cm$^{-1}$ (the value needed to observe the 2ZO mode)
and, still, we are considering a very regular graphene/Cr surface.

We remind that the defect-activated D peak observed in the Raman spectrum of graphene at $\sim$1350 cm$^{-1}$
can be activated by the presence of defects with very different nature~\cite{malard09, pimenta07, ferrari13}.
Thus, one could argue that the symmetry breaking induced by the presence of Cr atoms could possibly
also activate the D peak.
However, the Cr-induced symmetry breaking is substantially less disruptive
with respect to that induced, {\it e.g.}, by electron bombardment (which can create actual holes 
within the graphene sheet).
It is, thus, not surprising that the D peak observed in the presence of the 2ZO peak 
(Cr-coated not-annealed spectrum in Fig.~\ref{fig1}) is much less intense than the one
observed after ion bombardment (Fig.~\ref{fig4}).

%Finally, notice that the measured broadening of the G peak that is observed in the presence of the $U$ peak
%is $\sim$18~cm$^{-1}$ (FWHM), which is broader by $\sim$6~cm$^{-1}$ than the intrinsic
%FWHM of the G peak~\cite{bonini07}. 
%Within the present picture, this broadening is explained as the result of two contributions.
%First, the E$_{2g}$ is possibly split by a few cm$^{-1}$ and this will result in an apparent broadening of the G peak.
%Second, we also expect that the anharmonic coupling between the E$_{2g}$ and the 2ZO modes
%will induce a further broadening of the E$_{2g}$ which, however, cannot be properly described
%in the absence of further experimental hints concerning the atomic structure at the graphene/Cr interface.

\section {Conclusions}

In presence of Cr nanoparticles, the Raman spectrum of graphene exhibits
a new remarkable peak with a frequency of 1608~cm$^{-1}$ and intensity comparable
to that of the most prominent graphene Raman features.
Through direct experimental evidence we were able to dismiss several possible
mechanism at the origin of the 1608 cm$^{-1}$ peak. In particular,
the sample mechanical deformation and electronic doping that were deduced from Raman hyperspectral imaging are not unusual and thus cannot be at the origin of the new peak.
Moreover,
by intentionally generating defects on the sample through electron bombardment, we were able to show that the new peak is clearly distinct from the well-known defect-activated D$'$ peak at 1620~cm$^{-1}$(double resonance of a small wavevector phonon of the LO branch).
If we then consider the huge mount of Raman experiment published on graphene and
different experimental conditions exploited so far, it is unlikely that the new peak
can be attributed to activation mechanisms already documented for graphene.

We argue that the interaction of Cr with graphene softens the ZO phonon mode, downshifting its frequency $\omega_{\rm ZO}$, in such a way that the overtone frequency $2\omega_{\rm ZO}\sim\omega_{\rm G}$, where $\omega_{\rm G}$ is the frequency of the Raman active E$_{\rm 2g}$ mode. The new peak then becomes Raman-active within the mechanism known as Fermi resonance.
First-principles calculations of the $\omega_{\rm ZO}$ softening
on the graphene/Cr interface are compatible with this scenario,
also confirming that the analogous phenomenon should not be observable with other,
previously studied, metals such as nickel or copper due to the different kind of interaction with graphene.
According to calculations, the presence of Cr by itself is, possibly, also responsible for the mild
anharmonic mixing required for the 2ZO Raman activation.

Concluding, the present work provides strong hints for a Fermi resonance phenomenon, so far observed mainly in molecules and molecular crystals, in the case of a truly dispersive crystal. This is made possible by the unique nature of graphene, which allows the tuning of the ZO phonon frequencies through interaction with foreign species.

\section {Aknowledgments}

Calculations were done with the HPC resources of IDRIS (GENCI) within the allocation A0010910089.
The authors acknowledge financial support by ANR TRICO (ANR-11-NANO-025), Diracformag (ANR-14-CE32-0003), Organisio (ANR-15-CE09-0017), the European Union H2020 program under the grant 696656 Graphene Flagship and LANEF framework (ANR-10-LABX-51-01).

\section {Appendix}

\subsection {DFT computational details}
\label{appendix1}
We used ultra-soft pseudopotentials~\cite{vanderbilt90} and
spin polarization for Ni and Cr (in this last case it turns out, however, unnecessary).
Sampling of the electronic Brillouin zone was done with the k-point grid described below,
with Gaussian smearing of 0.01 Ry.
Electronic wavefunctions were expanded in plane-waves up to an energy cut-off, $\epsilon_{cut}$,
of 30 Ry (for Ni and Cu) and 60 Ry for Cr, the cut-off on the charge density being 12$\times\epsilon_{cut}$.
Phonon frequencies were determined within the approach of Ref.~\onlinecite{baroni01}.

Local density approximation~\cite{ceperley80,perdew81} (LDA) was used. Indeed,
it has been stressed that an accurate
description of equilibrium graphene-metal distances should include
van der Waals interactions~\cite{voloshina12}, which are poorly
described by standard DFT exchange-correlation functionals (LDA or GGA).
New approaches have been proposed to take into account dispersion interaction~\cite{vanin10}.
However, so far, most of the theoretical studies about graphene/metal interfaces
(see for instance Refs.~\onlinecite{xu10,khomyakov09,allard10}) have demonstrated
that LDA functionals, due to error cancellation, can predict both graphene-metal
distances and binding energies in fair agreement with experiments.

Bulk $fcc$ surfaces were simulated within periodic slab geometries of 6 atomic layers.
Graphene/$fcc$ results were obtained in the so called $top-fcc$~\cite{chanier15} configuration
which (according to both literature and present calculations) is the most stable.
For the on-$fcc$ structures, we used  32$\times$32 electronic k-point grids 
and 20 \AA~vacuum layer.
We tested that, for graphene/Cr systems, the structural geometry and $\omega_{\rm ZO}$ are not affected by
reducing the grid to 18$\times$18 and the vacuum to 10 \AA.
For the larger $bcc$ structure (Fig.~\ref{fig_structure}) we used a 2$\times$16 (shifted from the origin) grid
(which has a similar density as the 18$\times$18 for the $fcc$) and 10 \AA ~vacuum.
Atomic positions were relaxed to equilibrium by energy minimization
by fixing the in-plane lattice spacing and the positions of the lowest metallic layer.

Calculations on graphene/$fcc$ structures were done by imposing as in-plane lattice spacing
that calculated for the three bulk metals Cu, Cr, and Ni.
This implies that graphene lattice spacing is expanded by +2.1\%, +2.9\%, and -0.98\%,
respectively, with respect to the free-standing condition.
By repeating the calculations at the graphene equilibrium lattice spacing, the $\omega_{\rm ZO}$'s
change by $\lesssim3$~cm$^{-1}$.
Concerning the graphene/Cr$^{bcc}$ structure, we note that by imposing the in-plane equilibrium
lattice spacing of $bcc$ Cr in the lowest panel of Fig.~\ref{fig_structure},
graphene lattice spacing would be contracted by 3\% and 1\% along the horizontal and vertical directions, respectively.
We found preferable to avoid this anisotropy
(these calculations were not done to reproduce that particular structure
but to provide indications for a generic graphene/Cr interface)
and calculations were performed at the lattice spacings leaving graphene at equilibrium
in both directions (at the expense of a slight anisotropic  expansion of the underlying Cr).
The calculated equilibrium lattice spacings in \AA~of the bulk metals are:
$a^{\rm Cu}_{fcc}$=3.54 (2.50); $a^{\rm Cr}_{fcc}$=3.57 (2.52); $a^{\rm Ni}_{fcc}$=3.43 (2.43);
and $a^{\rm Cr}_{bcc}$=2.80 (2.43), the values in parenthesis being the nearest-neighbor atomic distance.
Calculated lattice spacing of graphene is $a^{Gr}$=2.45 \AA.

Finally, graphene/Cr$^{bcc}$ calculations are done by using one flat Cr layer, and the atomic
positions remain fixed during the relaxation of C atoms.
In the actual graphene/Cr$^{bcc}$  interface, Cr atoms at the surface are not exactly planar.
To estimate how this affects the present discussion,
we fully relaxed a graphene/Cr$^{bcc}$ slab containing 4 Cr layers.
The resulting vertical displacements of the Cr atoms of the outermost layer is not larger than 0.04 \AA~
(from the average flat position).
Then, we considered only the outermost Cr layer, thus relaxed, and let it interact with
the Gr layer. The resulting $\omega_{\rm ZO}$ is slightly higher ($\sim$1 cm$^{-1}$) than the one obtained from a flat Cr plane.
We, thus, consider this effect as negligible.

\subsection {Resonance model details}
\label{appendix2}

Let us consider Eq.~\ref{eq1} and the vibrational subspace ${\cal V}$ generated by
$|G_\alpha\rangle=a^\dagger_\alpha|0\rangle$ and
$|{\bf q}\rangle=\sqrt{N'/S'_0}b^\dagger_{{\bf q}}b^\dagger_{{\bf -q}}|0\rangle$,
with {\bf q} belonging to half of the BZ, $S'_0$ being
the surface of one half of the graphene BZ.
$|{\bf q}\rangle$ represents the excitation of two ZO phonons with opposite momentum.
The normalization is chosen so that $\langle{\bf q}|{\bf q'}\rangle=\delta^2({\bf q}-{\bf q'})$, where
$\delta^2$ is the two-dimensional Dirac delta distribution.
To simplify the problem, we restrict ${\cal V}$ to a half-circle centered at the BZ origin,
having radius $\overline{q}$ and we assume that $\omega_{\bf q}$ depends only on $|{\bf q}|$.
The integrations on the two-dimensional {\bf q} vectors can be
decomposed into radial, $q$=$|{\bf q}|$, and angular, $\theta$, components.
We then introduce the angular functions
$F_1(\theta)=1$,
$F_{2n}(\theta)=\sqrt{2}\cos(2n\theta)$, 
$F_{2n+1}(\theta)=\sqrt{2}\sin(2n\theta)$,
with  $n\in\mathbb {N}^*$.
These have the properties: $F_n(\theta+\pi)=F_n(\theta)$, $\int_0^\pi d\theta/\pi F_n(\theta)F_m(\theta)=\delta_{n,m}$,
and $\sum_n F_n(\theta)F_n(\theta')=\pi\delta(\theta-\theta')$ for $|\theta-\theta'|<\pi$.
We then define
\begin{equation}
|q,n\rangle=\sqrt{\pi q \overline{q}}\int_0^\pi\frac{d\theta}{\pi}F_n(\theta)|{\bf q}\rangle,
\end{equation}
where $q$ and $\theta$ are the radial and angular components of {\bf q}.
The angular decomposition is done on functions with periodicity $\pi$ because $|{\bf q}\rangle$ and
$|{\bf -q}\rangle$ represent the same excitation.
The normalization is chosen so that
$\langle q,n|q',m\rangle=\overline{q}\delta_{n,m}\delta(q-q')$, where $\delta(q)$ is the
one-dimensional Dirac delta distribution.

The vectors $|G_\alpha\rangle$ and $|q,n\rangle$ are then a basis of ${\cal V}$ and
the matrix elements of ${\cal H}$ within ${\cal V}$ are (consider $\hbar =1$):
\begin{eqnarray}
\langle G_\alpha |{\cal H}|G_{\alpha'} \rangle &=& \delta_{\alpha,\alpha'} \omega_{\rm G} \nonumber \\
\langle q,n |{\cal H}|q',m \rangle &=& 2\omega_q \overline{q} \delta_{n,m} \delta({q-q'}) \nonumber \\
\langle G_\alpha |{\cal H}|q,n\rangle &=& \sqrt{\frac{\pi q \overline{q}}{S'_0}} V_{\alpha,n}(q) \nonumber \\
&&V_{\alpha,n}(q) = \int_0^\pi V_\alpha({\bf q}) F_n(\theta) d\theta/\pi. \nonumber \\
\end{eqnarray}
A generic vector of ${\cal V}$ can be written as
\begin{eqnarray}
|\psi\rangle&=&\sum_\alpha x_\alpha|G_\alpha\rangle +\sum_n\int_0^{\overline {q}} \frac{q}{\overline{q}} \tilde{y}_{q,n}|q,n\rangle dq= \nonumber\\
&=&\sum_\alpha x_\alpha |G_\alpha\rangle + \frac{1}{\sqrt{N}} \sum_{n,i} \frac{q_i}{\overline{q}} y_{i,n} |q_i,n\rangle,
\end{eqnarray}
where, for computational purposes, in the second line the integral has been discretized
by summing on a uniform array of $N$ $q_i$ points ($i=1,N$) in 
the interval $[0,\overline{q}]$.
The solution of the eigenvalue problem ${\cal H}|\psi\rangle=\lambda|\psi\rangle$ is then reduced to solving the
linear system
%\begin{equation}
\begin{gather}
\left\{
\begin{split}
\omega_{\rm G}x_\alpha+\sum_{i,n} W_{\alpha,i,n} y_{i,n} &= \lambda x_\alpha \\
\sum_\alpha W_{\alpha,i,n} x_\alpha + 2 \omega_{q_i} y_{i,n} &= \lambda y_{i,n}, \\
\end{split} 
\right.
\\
{\rm with }~~~ W_{\alpha,i,n} = \sqrt{\frac{2\pi q_i\overline{q}}{NS'_0}} V_{\alpha,n}(q_i). ~~~~~~~~~\nonumber
\end{gather}
%\end{equation}
The unknowns are $\lambda$, $x_1$, $x_2$, and $\{y_{i,n}; i=1,...,N; n=1, ..., \infty\}$
and the problem reduces to the diagonalisation of a symmetric matrix.
With the present parameterizations for $V({\bf q})$ the problem is further simplified.
Indeed, if we consider the form of Eq.~(\ref{eq_p1}), the eigenvalues with component
$x_1\neq0$ are coupled only to the $\{y_{i,2}; i=1,...,N\}$ 
since $W_{1,i,n}=0$ for $n\neq 2$.
On the other hand, the eigenvalues with component
$x_2\neq0$ are coupled only to the $\{y_{i,3}; i=1,...,N\}$ 
since $W_{2,i,n}=0$ for $n\neq 3$).
If we now consider the form of Eq.~(\ref{eq_p2}), the eigenvalues with component
$x_1\neq0$ are coupled only to the $\{y_{i,1}; i=1,...,N\}$ 
since $W_{1,i,n}=0$ for $n\neq 1$,
while the eigenvalues with component $x_2\neq0$ are not coupled to the $\{y_{i,n}\}$.
Thus, in all the cases the matrix is block diagonal and the blocks in which one of the $x_\alpha$
is coupled to the $y_{i,n}$ variables have only $(N+1)\times(N+1)$ dimensions.
For Eq.~(\ref{eq_p1}) the two blocks associated with the two different polarizations
of the E$_{2g}$ phonon mode ($\alpha=1,2$) are identical. Thus, the two modes remain
degenerate.
For Eq.~(\ref{eq_p2}) one of the two polarization ($\alpha=1$) is coupled to the ZO's and the other ($\alpha=2$) is not.
Thus, while the eigenvalue associated with $\alpha=1$ shifts,
the one associated with $\alpha=2$ remains equal to $\omega_{\rm G}$.

Within the Lanczos procedure, the $(N+1)\times(N+1)$ matrix is reduced
to a tridiagonal form and this allows to write in a continued fraction expansion 
the matrix elements of the Green function associated to $\chi_{G_\alpha}(\omega)$,
see, e.g. Sect. V.8 of Ref.~\onlinecite{grosso}.


\begin{thebibliography}{99}
\bibitem{nota01}
Chromium coatings and their surface-passivating oxide
have long been used to protect oxidation-sensitive materials.
This is why they are deposited on Cu foils.
\bibitem{nota02}
Cr might form an alloy with Cu, if kinetic
barriers involved in this process are low-enough, or even a carbide in
presence of graphene~\cite{devincent92}. The resulting
nanoparticles, which are located underneath graphene, will in any case
be protected against oxidation by graphene itself
~\cite{chen11}. Pure Cr, a Cr-Cu alloy, and a Cr carbide are all
metallic, which is the key point here.
\end{thebibliography}
\end{document}